\documentclass[a4paper]{jpconf}
\usepackage{graphicx}
\usepackage{graphicx}
\usepackage{hyperref}
\usepackage{amsmath,amssymb,amsfonts}
\usepackage{algpseudocode}

\begin{document}

\title{Identification of a current-carrying subset of a percolation cluster using a modified wall follower algorithm}

\author{Renat~K~Akhunzhanov, Andrei~V~Eserkepov, and Yuri~Y~Tarasevich}
\ead{tarasevich@asu.edu.ru}

\address{Laboratory of Mathematical Modeling, Astrakhan State University, Astrakhan 414056, Russia}

\begin{abstract}
We have proposed and implemented a modification of the well-known wall follower algorithm to identify a backbone (a current-carrying part) of the percolation cluster. The advantage of the modified algorithm is identification of the whole backbone without visiting all edges. The algorithm has been applied to backbone identification in networks produced by random deposition of conductive sticks onto an insulating substrate. We have found that (i) for concentrations of sticks above the percolation threshold, the  strength of the percolating cluster quickly approaches unity; (ii) simultaneously, the percolation cluster is identical to its backbone plus simplest dead ends, i.e., edges that are incident to vertices of unit degree.
\end{abstract}

\section{Introduction\label{sec:intro}}

An approach describing the composites is the percolation theory~\cite{Stauffer}.
Percolation, i.e., the emergence of a connected subset (a cluster) that spans opposite boundaries in a disordered medium, has attracted the attention of the scientific community for several decades~\cite{Stauffer,Sahimi1994,BollobasRiordan2006,Grimmett1999,Kesten1982}. The occurrence of a percolation cluster drastically changes the physical properties of the medium, e.g., an insulator--conductor phase transition can be observed when the disordered medium is a mixture of conductive and insulating substances. However, only a fraction of the percolation cluster takes a part in the electrical conductivity~\cite{Skal1975SPS,DeGennes1976}. When the percolation cluster is treated as a random resistor network (RRN), a set of current-carrying bonds of such RRN is called the (effective) backbone~\cite{Herrmann1984JPhA}. The rest of the percolation cluster is  dead ends~\cite{Herrmann1984JPhA} (tag ends~\cite{Kirpatrick1978AIPCP}, tangling ends~\cite{Grassberger1992JPhA}) and  so-called perfectly balanced bonds (Wheatstone bridges). The electrical current through a perfectly balanced bond is absent because potential difference between its ends is equal to zero~\cite{Li2007JPhA}. The geometrical backbone is the union of all the self-avoiding walks (SAWs) between the two given points~\cite{Shlifer1979JPhysA}. SAW or a simple path is a path that contains no vertex twice. An algorithm for finding simple paths in a graph is based on depth-first search~\cite{Hopcroft1973CACM}. The geometrical backbone is the  effective one plus the ideally balanced bonds. Thus, the effective backbone is defined as the set of bonds that carry a current, while the geometrical backbone is the set of bonds that either carry a current, or are perfectly balanced~\cite{Batrouni1988PRA}. Another definition of the backbone says that it is the largest biconnected component of the graph~\cite{Kirpatrick1978AIPCP}. Such the definition may be confusing since a set of vertices is said to be biconnected, if each pair of vertices can be linked by at least two distinct paths. Hence, the two vertices connected by only one SAW can form no backbone according this definition. Although this definition is true when periodic boundary conditions (PBCs) are applied to the plane, i.e., a percolation on torus is considered, an examination of the electrical conductivity on a torus looks somewhat artificial.

Some of the bonds belonging to the backbone may carry the total current. These bonds are called red bonds or singly connected bonds; when they are cut, the current flow stops~\cite{Bunde1991percolationI}.

There are two different approaches for identification of backbones. On the one hand, one can use Ohm's law or Kirchhoff's rules to calculate potentials and currents in the RRN~\cite{Kirkpatrick1971PRL,Kirkpatrick1973RMP,Li2007JPhA,Kim2020JCPC}. However, direct calculations of electrical potentials and currents are based on floating-point arithmetic and, hence, produce round-off errors.  Due to these round-off errors, some ghost currents may arise which impedes the backbone extraction. Moreover, these calculations deal with huge systems of linear equations and require a lot of computer memory. Only relative small systems can be treated in these approaches because number of equations to be solved is proportional to the square of linear size of the system under consideration.

On the other hand, one can apply search algorithms on graphs~\cite{Tarjan1972SIAM,Roux1987JPhA,Moukarzel1998IJMPhC,Herrmann1984JPhA,Herrmann1984PRL,Grassberger1992JPhA,Mastorakos1993PRE,Porto1997PRE,Babalievski1998IJMPC,Sheppard1999JPhysA,Yin2000PhysB,Yin2003IJMPC,Trobec2017AES}.
In fact, some of the algorithms of backbone identification belong to maze solving algorithms (such as ``Ariadne's clew algorithm''~\cite{Mazer1998ACA}), which, in particular, are applied to wire routing on chips~\cite{Fattah2015NOCS}. Graph theory algorithms are sometimes difficult to understand or/and to realize. Some of them require storing not only original network but its dual~\cite{Roux1987JPhA}. Moreover, some algorithms can produce stack overflow because of recursion. All of the available graph-based algorithms remain storage limited, as some information at each node of the graph remains necessary~\cite{Alava2001}. In fact, application of these algorithms also is restricted to the RRN of moderate size.

Each of these two approaches have both advantages and disadvantages. A comparison and analysis of the algorithms devoted to identification of current-carrying part of the RRN can be found in Ref.~\cite{Tarasevich2018JPhCSbackbone}.

In this conference paper, we present a modification of a wall follower algorithm for a maze solving. The rest of the paper is constructed as follows. Section~\ref{sec:methods} describes some technical details of simulation
and our modification of the wall follower algorithm that extracts a geometrical backbone of a percolating cluster if any.
Section~\ref{sec:results} presents our main findings. Section~\ref{sec:concl} summarizes the main results.

\section{Modified wall follower algorithm\label{sec:methods}}

\begin{figure}[htb]
\begin{minipage}[c]{0.4\textwidth}
  \centering
\includegraphics[width=\textwidth]{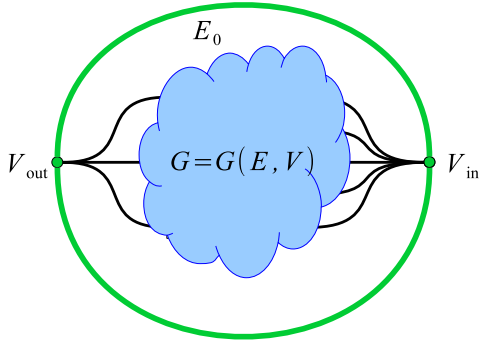}\\
\end{minipage}
\hfill
\begin{minipage}[c]{0.55\textwidth}
\caption{Transformation of the system under consideration for application of the modified wall follower algorithm.\label{fig:algorithm-figs}}
\end{minipage}
\end{figure}

Consider an embedding of an undirected simple planar graph $G = G(E,V)$ in two-dimensional Euclidean space $\mathbb{R}^2$.  We are looking for all simple paths (self-avoiding walks, SAWs) between ``entry'' vertex, $V_\text{in}$, and ``exit'' one, $V_\text{out}$. Three kinds of edges and vertices are distinguished in the algorithm. Initially, all edges and vertices are supposed to be ``black''. During the execution of the algorithm, the edges and vertices are colored in yellow or  green. Namely, ``green'' ones are classified to be a part of the geometric backbone, ``yellow'' ones are classified to be dead ends.
Before the algorithm starts working, the two ``green'' ghost edges should be added to the graph in the way depicted in Fig.~\ref{fig:algorithm-figs}. After identification of all SAWs between $V_\text{in}$ and $V_\text{out}$, the two ghost edges have to be removed.

\begin{figure}[!htb]
\centering
\includegraphics[width=\columnwidth]{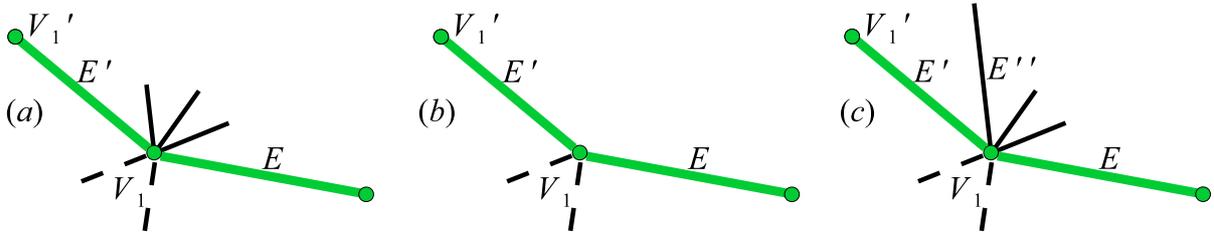}\\
\caption{(a) The edges incident to the vertex $V_1$. Edges between the edge $E$ and the next green edge $E'$ counterclockwise are under consideration (solid lines). Other edges (dashed lines) are not under consideration. (b) The case when $E$ is the nearest edge of any color clockwise of $E'$. (c)   The case when $E''$ is the nearest edge of any color clockwise of $E'$. \label{fig:MWFA}}
\end{figure}

Recursive procedure \texttt{NG}($E, V_1, V_2 , G$) is looking for SAWs between $V_1$ and $V_2$.
The procedure uses the two functions, viz., \texttt{NextEdge}($V, E, G, color, direction$) and \texttt{AdjacentVertex}($V, E, G$) (hereinafter, $V$ is a vertex of $G$ incident on edge $E$).
The function  \texttt{NextEdge}($V, E, G, color, direction$) returns an edge of the graph $G$ incident on the vertex $V$; the returned edge is the first one following the edge $E$ in the traversal direction indicated by parameter $direction$ and has the color indicated by the parameter $color$. The parameter $direction$ may be \texttt{clockwise} ($\circlearrowright$) or \texttt{counterclockwise} ($\circlearrowleft$). The parameter $color$ may be \texttt{green}, \texttt{yellow} or \texttt{any color}.
The function  \texttt{AdjacentVertex}($V, E, G$) returns the vertex of the graph $G$ that is incident to the edge $E$ and differs from the vertex $V$.
The procedure \texttt{NG} looks through all the edges incident to the vertex $V_1$, starting from the green edge $E$, and then counterclockwise, to the next green edge $E'$. Let us denote the next vertex of $E'$ as $V_1'$  (see Fig.~\ref{fig:MWFA}a). Then, starting from the green edge $E'$ clockwise, we are looking for the nearest edge of any color. If the nearest edge is $E$ (see Fig.~\ref{fig:MWFA}b), one should call \texttt{NG}($E', V'_1, V_2, G$). Otherwise (see Fig.~\ref{fig:MWFA}c), one should call lines 9--14 of the below pseudocode.

\begin{algorithmic}[1]
\Procedure{NG}{$E, V_1, V_2 , G$}
    \If{ $V_1 = V_2$}
        \Return
    \EndIf

    \State $E' \gets$ \Call{NextEdge}{$V_1,E,G, green, \circlearrowleft $}
    \State $V_1' \gets$ \Call{AdjacentVertex}{$E, V_1, G$}
     \State $E'' \gets$ \Call{NextEdge}{$V_1, E', G, any\ color,\circlearrowright$}
    \If{$ E'' = E$}
        \Call{NG}{$E', V_1', V_2 , G$}
    \Else
        \State $V_2' \gets$ \Call{WF}{$E, V_1, E'' , G$}
        \If{ $V_2' = V_1$}
            \Call{NG}{$E', V_1', V_2 , G$}
        \Else
           \State  \Call{NG}{$E', V_1', V_2' , G$}
          \State   \Call{NG}{$E, V_1, V_2 , G$}
        \EndIf
    \EndIf
\EndProcedure
\end{algorithmic}
To find all SAWs between $V_\text{in}$ and $V_\text{out}$, one should call the procedure as \texttt{NG}($E_0, V_\text{in},V_\text{out}, G$).

A modified wall follower algorithm is presented as a pseudocode of the function \texttt{WF}.
\begin{algorithmic}[1]
\Function{WF}{$E_1, V, E_2 , G$}
    \State $V.color\gets black$
    \State $V' \gets V$
    \State $E'_2 \gets E_2$
    \Loop
        \State $E'_2.color \gets yellow$
        \State $V' \gets$ \Call{AdjacentVertex}{$E'_2, V', G$}
        \If{$V'.color = green$}
            \State $V'' \gets V'$
            \Loop
                \State $E'_2.color \gets green$
                \State $V' \gets$ \Call{AdjacentVertex}{$E'_2,V',G$}
                \State $V'.color \gets green$
                \If{$V' = V$}
                    \Return{$V''$}
                \EndIf
                \State $E'_2 \gets$
                 \Call{NextEdge}{$V', E'_2, G, yellow,\circlearrowright$}
            \EndLoop
        \EndIf
         \State $E'_2 \gets$ \Call{NextEdge}{$V',E'_2,G, any\ color, \circlearrowleft$}
        \If{ $E'_2 = E_1$}
            \State $V.color \gets green$
            \State \Return{$V$}
        \EndIf
    \EndLoop
    \EndFunction
\end{algorithmic}

Zero-width sticks of unit length were randomly deposited onto a substrate of size $L \times L$ with PBCs until the desired number density was reached. Their centers are assumed to be independent and identically distributed (i.i.d.) on the substrate, while their orientations are assumed to be equiprobable. Hence, a homogeneous and isotropic network is produced. For basic computations, we used the system of size $L=32$. Finite-size effect has been additionally tested via variation of the system size.

Consider an arbitrary 2D network produced by random isotropic deposition of equally-sized zero-width sticks. Each stick is treated as a resistor with a specified electrical conductivity, $\sigma$. When this network is a subject to a potential difference, there are two natural possibilities~\cite{Redner2009}, viz.,
\begin{itemize}
  \item the ``bus-bar geometry'', when two parallel (super)conducting bars (buses) are attached to the opposite borders of the network; a potential difference (say, $V$ and 0, $V > 0$) is applied to these buses~\cite{Roux1987JPhA,Grassberger1992JPhA,Trobec2017AES},
  \item the ``two-point geometry'', when a potential difference is applied to the two distinct sites, so that an electrical current, $I$, injected into one site (source) and the same current withdrawn from the other (sink)~\cite{Herrmann1984JPhA}.
\end{itemize}
In the case of superconducting buses, ``bus-bar geometry'' can be turn to ``two-point geometry'' by addition of ghost vertices.

To detect the spanning cluster, the Union–Find algorithm~\cite{Newman2000PRL,Newman2001PRE} modified for continuous systems~\cite{Li2009PRE,Mertens2012PRE} was applied. When the spanning cluster was found, all other clusters were removed since they cannot contribute in the electrical conductivity. All edges of the spanning cluster incident on vertex of unit valence were cut off, since, obviously, they are simplest dead ends. According to Ref.~\cite{Kumar2017JAP}, we denote such the preprocessed spanning cluster as ``approximate backbone''. To detect the backbone of the spanning cluster, the modified wall follower algorithm was used.
When the geometrical backbone has been identified, an adjacency matrix was formed for it. With this adjacency matrix in hands,  Kirchhoff's current law was used for each junction of sticks, and Ohm’s law for each circuit between two junctions.

The computer experiments were repeated 100 times. The error bars in the figures correspond to the standard deviation of the mean. When not shown explicitly, they are of the order of the marker size.

A particular case of a planar graph is $N$ zero-width sticks of length $l$ which centers are assumed to be i.i.d. within a square domain $\mathcal{D}$ of size $L \times L$ with periodic boundary conditions ($\mathcal{D} \in \mathbb{R}^2$), i.e., $x,y \in [0;L]$, where $(x,y)$ are the coordinates of the center of the stick. The relation $L>l$ is assumed. All intersections of sticks with the lines $x=L$ and $x=0$ are supposed to be vertices (``entries'' and ``exits'', respectively).
To apply the above algorithm, we transform a bus-bar geometry to a two-point geometry by adding two ghost vertices, viz., $V_\text{in}$ is adjacent to all vertices belonging to ``entries'', while  $V_\text{out}$ is adjacent to all vertices belonging to ``exits''. 
In such a way, the problem of geometrical backbone identification for bus-bar geometry is transforms into the one for two-point geometry.

In many cases, modified wall follower algorithm can identify the backbone without visiting all edges. In a graph produced by random isotropic deposition of zero-width sticks onto a plane, near the percolation threshold, a fraction of unvisited edges after complete identification of the backbone approaches to 0.5.

\section{Results\label{sec:results}}
\begin{figure}[htb]
\begin{minipage}[c]{0.46\textwidth}
  \centering
\includegraphics[width=\columnwidth]{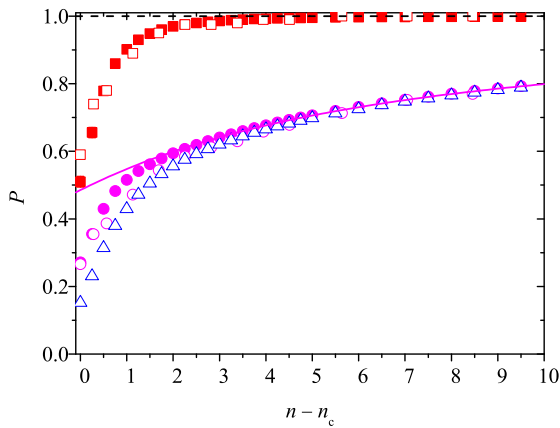}\\
\end{minipage}
\hfill
\begin{minipage}[c]{0.49\textwidth}
\caption{Dependencies of the strength of the percolation cluster (\fullsquare\ and \opensquare),  of the strength of the approximate backbone (\fullcircle\ and \opencircle), and of the strength of the backbone (\opentriangle) on the shifted  number density, $n-n_\text{c}$. Full symbols correspond to our results, while the open symbols represents the results extracted from Ref.~\cite{Kumar2017JAP}. Solid line (\full) corresponds to analytical estimation from Ref.~\cite{Kumar2017JAP}.\label{fig:fractions}}
\end{minipage}
\end{figure}

Figure~\ref{fig:fractions} demonstrate how the quantities of interest depends on the shifted number density, $n-n_\text{c}$. Solid symbols present our results, while the open symbols represents the results extracted from Ref.~\cite{Kumar2017JAP}. The strength of the percolation cluster approaches unit reflecting the fact that almost all sticks belong to the percolation cluster when $n \gtrapprox 2n_\text{c}$. This observation is quite consistent with the previously published results~\cite{Kumar2017JAP}. At the large number density, the backbone and the approximate backbone~\cite{Kumar2017JAP} are indistinguishable within simulation accuracy. This fact validates  the assumption~\cite{Kumar2017JAP} that the percolation cluster is identical to its geometrical backbone plus simplest dead ends, i.e., edges incident on the vertices of the unit degree.  Solid line corresponds to theoretical estimate of the approximate backbone offered in Ref.~\cite{Kumar2017JAP}.

\section{Conclusion\label{sec:concl}}
We have proposed and implemented a modified wall follower algorithm for backbone identification. The algorithm was applied to backbone identification for different system sizes and concentrations of conductive sticks.
We have found that (i) for concentrations of sticks above the percolation threshold, the  strength of the percolating cluster quickly approaches unity as concentration of sticks increase; (ii) simultaneously, the percolation cluster is almost identical to its backbone plus simplest dead ends, i.e., edges that are incident to the vertices of degree one.

\ack
Y.Y.T. and A.V.E. acknowledge the funding from the Foundation for the Advancement of Theoretical Physics and Mathematics ``BASIS'', grant~20-1-1-8-1. The authors would like to thank  I.I.Gordeev for stimulating discussions.

\section*{References}
\bibliographystyle{iopart-num}
\bibliography{backboneshort}

\providecommand{\newblock}{}
\begin{thebibliography}{10}
\expandafter\ifx\csname url\endcsname\relax
  \def\url#1{{\tt #1}}\fi
\expandafter\ifx\csname urlprefix\endcsname\relax\def\urlprefix{URL }\fi
\providecommand{\eprint}[2][]{\url{#2}}

\bibitem{Stauffer}
Stauffer D and Aharony A 1994 {\em Introduction to percolation theory\/} 2nd ed
  (London: Taylor \& Francis) ISBN 9781315274386

\bibitem{Sahimi1994}
Sahimi M 1994 {\em Applications of percolation theory\/} (Taylor \& Francis)
  ISBN 9780429080449

\bibitem{BollobasRiordan2006}
Bollob\'{a}s B and Riordan O 2006 {\em Percolation\/} (Cambridge: Cambridge
  University Press) ISBN 9781139167383

\bibitem{Grimmett1999}
Grimmett G~R 1999 {\em Percolation\/} (Berlin, Heidelberg: Springer-Verlag)
  ISBN 978-3-540-64902-1

\bibitem{Kesten1982}
Kesten H 1982 {\em Percolation theory for mathematicians\/} ({\em Progress in
  Probability and Statistics\/} vol~2) (Mass.: Birkh\"auser Boston) ISBN
  3-7643-3107-0

\bibitem{Skal1975SPS}
Skal A~S and Shklovskii B~I 1975 {\em Sov. Phys. Semicond.\/} {\bf 8}
  1029--1032 ISSN 0038-5700

\bibitem{DeGennes1976}
De~Gennes P~G 1976 {\em J. Physique Lett.\/} {\bf 37} 1--2 ISSN 0302-072X

\bibitem{Herrmann1984JPhA}
Herrmann H~J, Hong D~C and Stanley H~E 1984 {\em J. Phys. A: Math. Gen.\/} {\bf
  17} L261--L266 ISSN 1361-6447

\bibitem{Kirpatrick1978AIPCP}
Kirpatrick S 1978 {\em AIP Conf. Proc.\/} {\bf 40} 99--117 ISSN 0094-243X

\bibitem{Grassberger1992JPhA}
Grassberger P 1992 {\em J. Phys. A: Math. Gen.\/} {\bf 25} 5475--5484 ISSN
  1361-6447

\bibitem{Li2007JPhA}
Li C and Chou T~W 2007 {\em J. Phys. A: Math. Theor.\/} {\bf 40} 14679 ISSN
  1751-8121

\bibitem{Shlifer1979JPhysA}
Shlifer G, Klein W, Reynolds P~J and Stanley H~E 1979 {\em J. Phys. A: Math.
  Gen.\/} {\bf 12} L169--L174 ISSN 0305-4470

\bibitem{Hopcroft1973CACM}
Hopcroft J and Tarjan R 1973 {\em Commun. ACM\/} {\bf 16} 372--378 ISSN
  0001-0782

\bibitem{Batrouni1988PRA}
Batrouni G~G, Hansen A and Roux S 1988 {\em Phys. Rev. A\/} {\bf 38}(7)
  3820--3823 ISSN 2469-9926

\bibitem{Bunde1991percolationI}
Bunde A and Havlin S 1991 {\em Fractals and Disordered Systems\/} ed Bunde A
  and Havlin S (Berlin, Heidelberg: Springer Berlin Heidelberg) pp 51--96 ISBN
  978-3-642-51435-7

\bibitem{Kirkpatrick1971PRL}
Kirkpatrick S 1971 {\em Phys. Rev. Lett.\/} {\bf 27}(25) 1722--1725 ISSN
  0031-9007

\bibitem{Kirkpatrick1973RMP}
Kirkpatrick S 1973 {\em Rev. Mod. Phys.\/} {\bf 45} 574--588 ISSN 0034-6861

\bibitem{Kim2020JCPC}
Kim D and Nam J 2020 {\em J. Phys. Chem. C\/} {\bf 124} 986--996 ISSN 1932-7447

\bibitem{Tarjan1972SIAM}
Tarjan R 1972 {\em SIAM J. Comput.\/} {\bf 1} 146--160 ISSN 0097-5397

\bibitem{Roux1987JPhA}
Roux S and Hansen A 1987 {\em J. Phys. A: Math. Gen.\/} {\bf 20} L1281--L1285
  ISSN 1361-6447

\bibitem{Moukarzel1998IJMPhC}
Moukarzel C 1998 {\em Int. J. Mod. Phys. C\/} {\bf 9} 887--895 ISSN 0129-1831

\bibitem{Herrmann1984PRL}
Herrmann H~J and Stanley H~E 1984 {\em Phys. Rev. Lett.\/} {\bf 53}(12)
  1121--1124 ISSN 0031-9007

\bibitem{Mastorakos1993PRE}
Mastorakos J and Argyrakis P 1993 {\em Phys. Rev. E\/} {\bf 48}(6) 4847--4850
  ISSN 2470-0045

\bibitem{Porto1997PRE}
Porto M, Bunde A, Havlin S and Roman H~E 1997 {\em Phys. Rev. E\/} {\bf 56}(2)
  1667--1675 ISSN 2470-0045

\bibitem{Babalievski1998IJMPC}
Babalievski F 1998 {\em Int. J. Mod. Phys. C\/} {\bf 09} 43--60 ISSN 0129-1831

\bibitem{Sheppard1999JPhysA}
Sheppard A~P, Knackstedt M~A, Pinczewski W~V and Sahimi M 1999 {\em J. Phys. A:
  Math. Gen.\/} {\bf 32} L521--L529 ISSN 1361-6447

\bibitem{Yin2000PhysB}
Yin W~G and Tao R 2000 {\em Physica B\/} {\bf 279} 84--86 ISSN 0921-4526

\bibitem{Yin2003IJMPC}
Yin W~G and Tao R 2003 {\em Int. J. Mod. Phys. C\/} {\bf 14} 1427--1437 ISSN
  0129-1831

\bibitem{Trobec2017AES}
Trobec R and Stamatovic B 2017 {\em Adv. Eng. Softw.\/} {\bf 103} 38--45 ISSN
  0965-9978

\bibitem{Mazer1998ACA}
Mazer E, Ahuactzin J~M and Bessi{\`e}re P 1998 {\em J. Artif. Int. Res.\/} {\bf
  9} 295--316 ISSN 1076-9757

\bibitem{Fattah2015NOCS}
Fattah M, Airola A, Ausavarungnirun R, Mirzaei N, Liljeberg P, Plosila J,
  Mohammadi S, Pahikkala T, Mutlu O and Tenhunen H 2015 {\em Proc. 9th Int.
  Symp. on Networks-on-Chip\/} NOCS '15 (New York, NY, USA: ACM) pp 18:1--18:8
  ISBN 978-1-4503-3396-2

\bibitem{Alava2001}
Alava M~J, Duxbury P~M, Moukarzel C~F and Rieger H 2001 {\em Phase Transitions
  and Critical Phenomena\/} vol~18 ed Domb C and JL~Lebowitz J~L (Academic
  Press) chap Exact combinatorial algorithms: {Ground} states of disordered
  systems, pp 143--317 ISSN 1062-7901

\bibitem{Tarasevich2018JPhCSbackbone}
Tarasevich Y~Y, Burmistrov A~S, Goltseva V~A, Gordeev I~I, Serbin V~I, Sizova
  A~A, Vodolazskaya I~V and Zholobov D~A 2018 {\em J. Phys. Conf. Ser.\/} {\bf
  955} 012021 ISSN 1742-6596

\bibitem{Redner2009}
Redner S 2009 {\em Encyclopedia of Complexity and Systems Science\/} ed Meyers
  R~A (New York, NY: Springer New York) pp 3737--3754 ISBN 978-0-387-30440-3

\bibitem{Newman2000PRL}
Newman M~E~J and Ziff R~M 2000 {\em Phys. Rev. Lett.\/} {\bf 85}(19) 4104--4107
  ISSN 0031-9007

\bibitem{Newman2001PRE}
Newman M~E~J and Ziff R~M 2001 {\em Phys. Rev. E\/} {\bf 64}(1) 016706 ISSN
  2470-0045

\bibitem{Li2009PRE}
Li J and Zhang S~L 2009 {\em Phys. Rev. E\/} {\bf 80}(4) 040104 ISSN 2470-0045

\bibitem{Mertens2012PRE}
Mertens S and Moore C 2012 {\em Phys. Rev. E\/} {\bf 86}(6) 061109 ISSN
  2470-0045

\bibitem{Kumar2017JAP}
Kumar A, Vidhyadhiraja N~S and Kulkarni G~U 2017 {\em J. Appl. Phys.\/} {\bf
  122} 045101 ISSN 0021-8979

\end{thebibliography}

\end{document}